\documentclass[prl,twocolumn,showpacs,superscriptaddress,floatfix,footinbib,amsmath,amssymb]{revtex4}

\usepackage[utf8]{inputenc}
\usepackage{graphicx}% Include figure files
\usepackage[normalem]{ulem}
\usepackage[usenames,dvipsnames]{color}
\usepackage[nointegrals]{wasysym}
\usepackage{commands}
\usepackage{cleveref}
\usepackage{upgreek}

\begin{document}

%\preprint{APS/123-QED}

\title{Multistability and spin diffusion enhanced lifetimes in dynamic nuclear polarization in a double quantum dot}

\author{F. Forster}\affiliationMunich
\author{M. Mühlbacher}\affiliationMunich
\author{D. Schuh}\affiliationRegensburg
\author{W. Wegscheider}\affiliationETH
\author{G. Giedke}\affiliationMPQ\affiliationIKER\affiliationDIPC
\author{S. Ludwig}\affiliationMunich\affiliationPDI

\date{\today}

\begin{abstract}
The control of nuclear spins in quantum dots is essential to explore their many-body dynamics and exploit their prospects for quantum information processing. We present a unique combination of dynamic nuclear spin polarization and electric-dipole-induced spin resonance in an electrostatically defined double quantum dot (DQD) exposed to the strongly inhomogeneous field of two on-chip nanomagnets. Our experiments provide direct and unrivaled access to the nuclear spin polarization distribution and allow us to establish and characterize multiple fixed points. Further, we demonstrate polarization of the DQD environment by nuclear spin diffusion which significantly stabilizes the nuclear spins inside the DQD.
\end{abstract}

\maketitle

In III-V semiconductors the weak hyperfine interaction between nuclear
and electron spins has a strong impact on the electron spin dynamics
owing to the fact that each conduction band electron interacts with a
large number of nuclei
\cite{Urbaszek2013,Chekhovich2013,Taylor2007,CoBa09}. This
situation can give rise to dynamic nuclear spin polarization (DNSP) \cite{Ono2004,Danon2009,Frolov2012,Petersen2013,Bracker2005,FAH07,Hoegele2012,InPl08,Gullans2010,NRH14} and exciting many-body quantum physics such as complex hysteretic dynamics \cite{Ono2004,Latta2009,Urbaszek2013}, multistabilities \cite{Vink2009,Danon2009}, collectively enhanced transport \cite{Schuetz2012}, and dissipative phase transitions \cite{Rudner2010b,Kessler2012}. However, the thermal fluctuations of nuclear spins, even present at cryogenic temperatures, also cause decoherence of spin qubits \cite{Fischer2009,Bluhm2011}. Ignoring correlations, the influence of the nuclei on the electron
 spin dynamics is usually described within a semiclassical mean field
 approach, which expresses the nuclear spin polarization in terms of
 an effective magnetic field for the electron spin, \Bnr, the Overhauser field \cite{Overhauser1953}. In a DQD charged by one electron in each dot, thermally fluctuating nuclear spins result in a field difference \Bnd\ in the order of few mT between the two dots \footnote{$\Bnr$ is the Overhauser field acting on an electron in the lowest orbital eigenstate of the QD centered at position $\mathbf r$. $\Bnd$ is the difference of \Bnr\ acting on two electrons, each located in one of the dots. \Bmd\ below is defined likewise.}. In equilibrium, the time averages of \Bnr\ and \Bnd\ vanish. This fluctuating field nevertheless causes a weak mixing of singlet and triplet states \cite{Koppens2005,Johnson2005b,Taylor2007} being explored for quantum information processing \cite{Bluhm2011,Ribeiro2010}. To control this mixing, it would be necessary to stabilize the difference of the effective magnetic fields in the two dots, $\Bnd$.

\begin{figure}
\includegraphics[width=1\columnwidth]{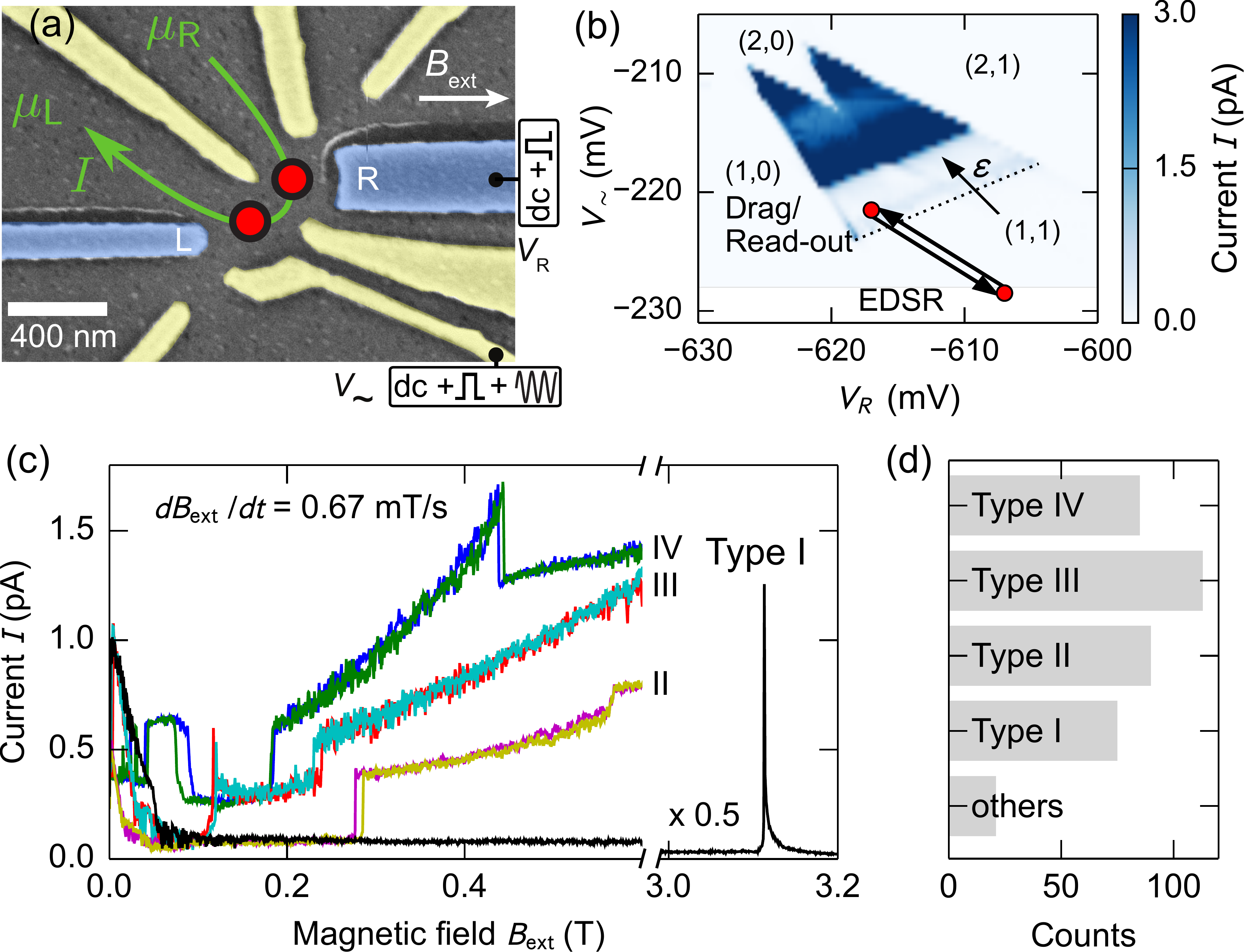}
\caption{
(a) Scanning electron microscope image of a DQD defined in a two-dimensional electron system 85\,nm beneath the surface of a GaAs/AlGaAs heterostructure \cite{Forster2015}: GaAs surface in gray, gold gates in yellow, cobalt gates in blue. Red circles indicate approximate quantum dot positions, green arrows the physical current direction for $\mur>\mul$, a white arrow the direction of \Bext.
(b) Charge stability diagram $I(V_\sim,V_\text R)$. An arrow marks the detuning axis, a dotted line $\varepsilon = 0$. A double arrow indicates pulsing of gate voltages $V_\sim$ and $V_\text R$ during an EDSR experiment [see labels in (a)]; the modulation $V_\sim=V_\sim^0+v\sin(\omega t)$ with $v\simeq3\,$mV is applied at the red point labeled EDSR.
(c) $I(\Bext)$ as $\Bext$ is swept. Variations suggest the existence of multiple FPs.
(d) Distribution of current traces $I(\Bext)$ corresponding to FPs I--IV and ``others'' (referring to traces which did not fit to types I--IV) detected within $384$ sweeps.
}
\label{fig:pol_types}
\end{figure}

Here, we combine DNSP with EDSR to study the dynamic polarization of nuclear spins on the one hand, and the decay of the polarization on the other hand. We demonstrate the existence of multiple attractive fixed points (FPs) in the steady-state solution of the driven system, where the decay of \Bnr\ is exactly canceled by its dynamical build up \cite{Danon2009,Rudner2011,Petersen2013}. Our results demonstrate that the FPs differ from each other by their spatial distributions of \Bnr. Relevant for spin qubit applications, the singlet-triplet mixing of each FP can thereby be fine tuned by adjusting external parameters such as the external field \Bext\ [direction as in \fig{fig:pol_types}(a)] or the energy detuning $\varepsilon$ between the two dots [i.e.\ the singlet configurations $(1,1)$ and $(2,0)$, where $(n,m)$ denotes the number of electrons in the left $(n)$ and right $(m)$ dot]. Attractive FPs in DNSP are often characterized by
a narrow nuclear spin distribution, hence sharply reduced nuclear spin fluctuations, which provides possible advantages for the preparation of coupled dots for quantum information applications \cite{Bluhm2010,Chekhovich2013}. Our measurements allow us to characterize FPs by their nuclear spin distribution, and their dynamic and static stability. In particular, we show that the diffusion of nuclear spins outside the DQD has a strong influence on the build-up and decay dynamics of their polarization inside the DQD which it further stabilizes.
Such an enhanced stability of FPs promises positive impact on the coherence of the electron and nuclear spin states.

Our DQD design, presented in \fig{fig:pol_types}(a), incorporates two single-domain nanomagnets. They generate an additional, static inhomogeneous field \Bmr, such that the total effective field difference is $\Delta\mathbf{\Beff}=\Bnd+\Bmd$. In equilibrium, the static $|\Bmd|\simeq45\,$mT \cite{Forster2015} exceeds the fluctuations of $\Bnd$ of $\sim2\,$mT \cite{Taylor2007} by far, degrading \Bnd\ to a weak perturbation. Important for qubit applications this stabilizes the singlet-triplet splitting and yields advantages in controlling the nuclear spin dynamics \cite{Petersen2013}. Furthermore, our sizable \Bmd\ causes a corresponding separation of the Zeeman energy in the two dots and allows us to resolve EDSR experiments in the individual dots \cite{Forster2015}.
We operate our DQD in the vicinity of the (1,1) $\leftrightarrow$ (2,0) charge transition. In response to the applied voltage $V=(\mur-\mul)/e=1\,$mV electrons tunnel one by one through the DQD via the cycle (1,0)$\rightarrow$(1,1)$\rightarrow$(2,0)$\rightarrow$(1,0). We measure the resulting current which is, however, strongly suppressed by Pauli spin blockade (PSB) of the transition (1,1)$\rightarrow$(2,0) \cite{Ono2002}. In the stability diagram of our DQD in \fig{fig:pol_types}(b), an extended region of PSB is clearly visible as reduced current at the base of the current-carrying double triangle. Details are explained in Ref.\ \onlinecite{Forster2015} for the identical sample.

We follow two complementary approaches to study the nuclear spin dynamics. First, we actively polarize the nuclear spins by sweeping $B_\text{ext}$ and driving electrons through the DQD. We measure the background leakage current still flowing in PSB, which grows with increasing singlet-triplet mixing being proportional to the components of \Beffd\ \cite{Petersen2013} (if we ignore weak influences of co-tunneling and spin-orbit interaction). In our second approach, we let \Bnr\ (produced as described above) decay and directly measure it by performing electric dipole induced spin resonance (EDSR). We have experimented with several scenarios, but for better comparability here we discuss only measurements taken under the following conditions: we start with an initialization time $t_\text{init}=180\,$s with $\Bext=V=0$ at $\varepsilon\simeq0.1\,$meV (red dot in \fig{fig:pol_types}(b) labeled "Drag"; the interdot tunnel coupling is tuned to $\tc\simeq20\,\upmu$eV) to let any remaining \Bnr\ decay. Next, we apply $V=1\,$mV (at otherwise identical settings) and sweep \Bext\ at the rate $0.67\,\text{mT}/$s to a finite value and then sweep back to zero at $-3.3\,\text{mT}/$s. To avoid complications by long time memory effects, before each measurement series we preconditioned the system with a number of identical sweeps.

Figure \ref{fig:pol_types}(c) shows $I(\Bext)$ during typical sweeps to $\Bextmax=0.6\,$T. Even though we keep dot and sweep parameters identical we find four different characteristic current traces each one occurring multiple times [\fig{fig:pol_types}(d)] but in arbitrary order. Within each type $I(\Bext)$ is reproducible, even including sudden current steps and the noise level, see \fig{fig:pol_types}(c). The different curve types $I(\Bext)$ are stable over many minutes (jumps between them occur very rarely at sizable \Bext, not more than in one out of 50 sweeps). They strongly depend on \Bext, and persist even for very slow sweeps, but are lost if the sweep is performed too fast. We will show below, that the magnetic field sweeps are accompanied by a build-up of sizable (and type-dependent) nuclear polarizations.

Knowing that $I$ scales with \Beffd, this all points to corresponding stable FPs with steady-state nuclear spin configurations $\mathbf B_\text{nuc}(\mathbf r,\Bext)$. 
Traces II-IV contain sudden jumps (within less than a second), which might indicate reproducible transitions between some FPs at specific values of \Bext. Before each sweep we let \Bnr\ decay by setting $V=0$, hence $I=0$, where DNSP is absent (and only the trivial FP in equilibrium with $\Bnr=0$ is left). Our experiments indicate that within the first few seconds of $I\ne0$ the nuclear spins arrange themselves at one of the FPs (even at $\Bext=0$). Figure \ref{fig:pol_types}(d) presents a statistics of the rate of different curve types. It suggests that for our settings four FPs are almost equally likely occupied under these conditions. Interestingly, the order at which the different types occur is random, likely being related to random fluctuations. In our case these could be thermal fluctuations of $\mathbf B_\text{nuc}(\mathbf r,\Bext)$ \cite{Reilly2008a,Chekhovich2013} or random telegraph noise in the local DQD potential, called charge noise \cite{Jung2004,Pioro-Ladriere2005,Taubert2008}.

Type I curves [see \fig{fig:pol_types}(c)] are characterized by a reduction of the current from $I\simeq 1\,$pA at $\Bext\simeq0$ to $I<80\,$fA at $\Bext>0.1\,$T. In Ref.\ \onlinecite{Petersen2013}, type I has been associated with the resonance between the T$_+$-triplet and the singlet state (both for one electron in each dot). In our system, the strong reduction of $I(\Bext)$ implies that especially the components of \Beffd\ perpendicular to \Bext\ should be reduced to $\Delta\Beff^\perp\simeq0$.  Likewise, the other FPs with larger currents correspond to more inhomogeneous \Bnr\ at sizable \Bext.

\begin{figure}
\includegraphics[width=1\columnwidth]{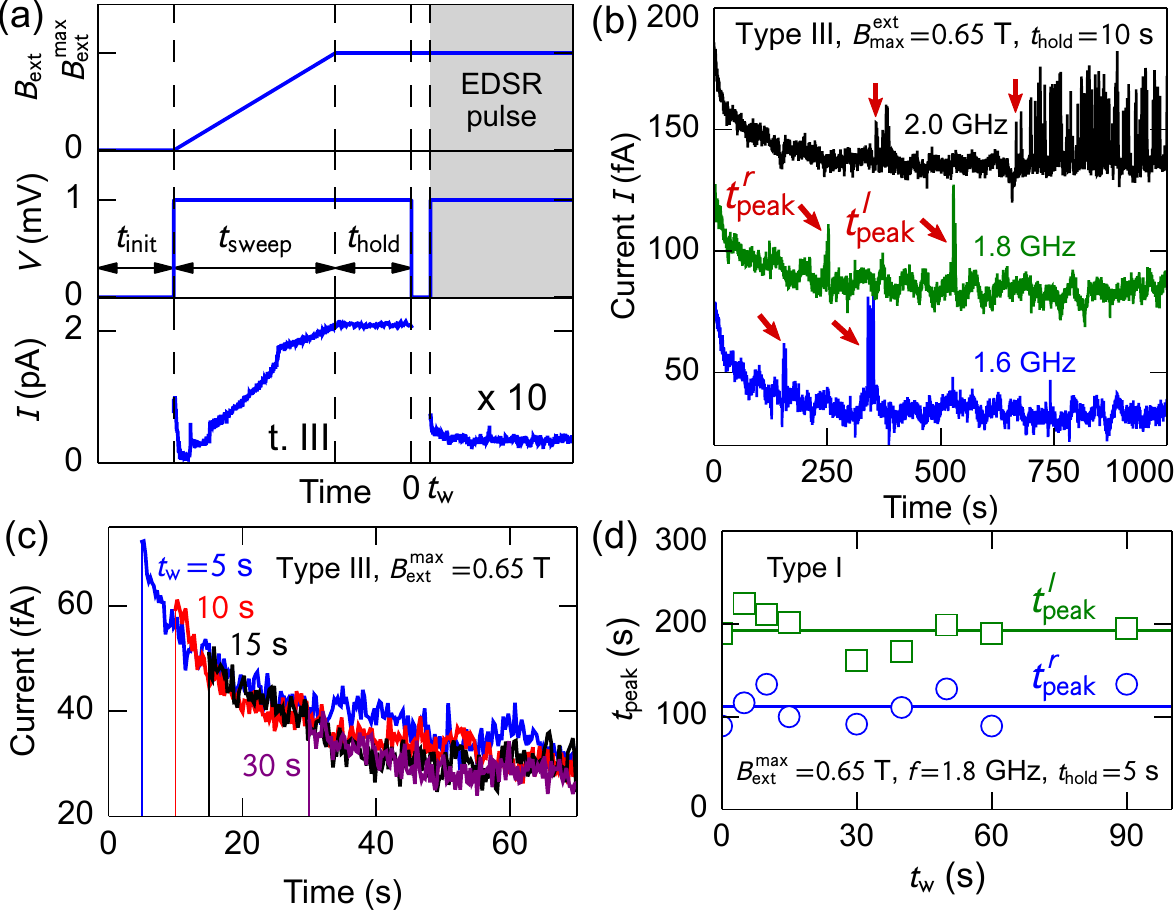}
\caption{ 
(a) Measuring scheme and an example current trace $I(t)$ (type III); $\Bnr$ begins to decay at $t=0$. (See main text for details.)
(b) $I(t)$ during the decay of \Bnr\ for various EDSR frequencies. Current peaks marked with red arrows indicate EDSR resonances. (For clarity the upper traces are vertically offset in steps of $50\,$fA.)
(c) Initial current decay for various $t_\text{w}$.
(d) EDSR resonance times \tr\ and \tl\ as function of $t_\text{w}$. Horizontal lines are mean values.
}
\label{fig:edsr}
\end{figure}
Utilizing EDSR experiments, we can measure the effective magnetic field value to monitor $\Bnr$. Here, we measure its decay in the two dots after we have built up \Bnr\ by DNSP. As sketched in \fig{fig:edsr}(a), we start by sweeping \Bext\ to \Bextmax\ within $t_\text{sweep}$ as described above. There, we hold $\Bext$ and $V $ constant during $t_\text{hold} = 120\,$s. For the remainder of the experiment we keep $\Bext=\Bextmax$.
Next, at $t=0$, we initiate the decay of \Bnr\ by switching $V$ off to stop DNSP. After the waiting period $t_\text w$, we go back to $V=1\,$mV. If $t_\text w$ was long enough, the FP is lost and \Bnr\ continues decaying. To measure this, we perform EDSR by periodically pulsing to $\varepsilon\simeq-0.5\,\text{meV}$ [see \fig{fig:pol_types}(b)] and applying an rf-modulation at a fixed frequency to the gate voltage $V_\sim$.
 Details of our EDSR procedure are explained in Ref.\   \onlinecite{Forster2015}. At $\varepsilon\ll-\tc$ the electrons, confined in the DQD and affected by EDSR, can be considered localized in the individual dots. As long as the rf-bursts are off-resonance, $I$ remains small, but whenever $hf=g\mB\Beff$ in one of the two dots, PSB is lifted and we expect enhanced current.

Typical traces $I(t)$ are plotted in \fig{fig:edsr}(b) for the case of type III curves for three different frequencies and $\Bextmax=0.65\,$T. The frequency independent gradual decrease of $I$ within the first 250\,s is related to the decay of \Bnr\ away from the FP. It indicates that the FP is close to a singlet-triplet resonance, where $I$ has a maximum. On top of this background, we observe current spikes indicating EDSR resonances [arrows in \fig{fig:edsr}(b)]. Interestingly, we find two distinct resonances per decay curve, the first occurring at \tr\ and the second at \tl, which we assign to the right and left dot, respectively (see below). A resonant EDSR experiment can also lead to DNSP (more so at higher frequencies) \cite{Laird2007,Kroner2008,Danon2009,Vink2009,Obata2012}. The uppermost $I(t)$ curve in \fig{fig:edsr}(b) measured at $f=2\,$GHz demonstrates this effect, where the second EDSR spike repeats multiple times and causes $I$ to fluctuate strongly. Important for our analysis, no DNSP is induced as long as $f$ is off-resonant. This is always the case in each dot before the corresponding current peak occurs the first time, marking the relevant \tl\ and \tr.

For measuring the undisturbed decay of \Bnr, it would be desirable to avoid any DNSP effects during the decay. As a continuous $V=0$ and no rf-modulation are no option, we carefully monitor and set up the experiments to avoid unwanted DNSP effects. In particular, we found that the background current does not cause any DNSP away from the FPs: for instance, variations of $t_\text w$ (at $V=0$) do not influence the initial decay of $I(t)$, see \fig{fig:edsr}{(c)}. In addition, \tr\ and \tl\ are independent of $t_\text w$ as demonstrated in \fig{fig:edsr}(d). However, a correlation is evident in the jitter between \tr\ and \tl\ (the correlation coefficient is 0.87). 
Such a behavior is expected if the origin of the fluctuations is related to noise occurring with frequencies small compared to $1/(\tl-\tr)$. This favors charge noise, which has its weight at long time scales \cite{Fujisawa2000,Jung2004,Pioro-Ladriere2005,Taubert2008}, over the thermal fluctuations of \Bnr, which are fast compared to $\tl-\tr$ \cite{Chekhovich2013}.
Charge noise modulates the geometry of the confinement potential. The details of the latter determine parameters with a possibly strong impact on the static and dynamic properties of the FPs, including the detuning $\varepsilon$, the positions of the charge centers, consequently \Bmd, and the decay of \Bnr. 

\begin{figure}
\includegraphics[width=1\columnwidth]{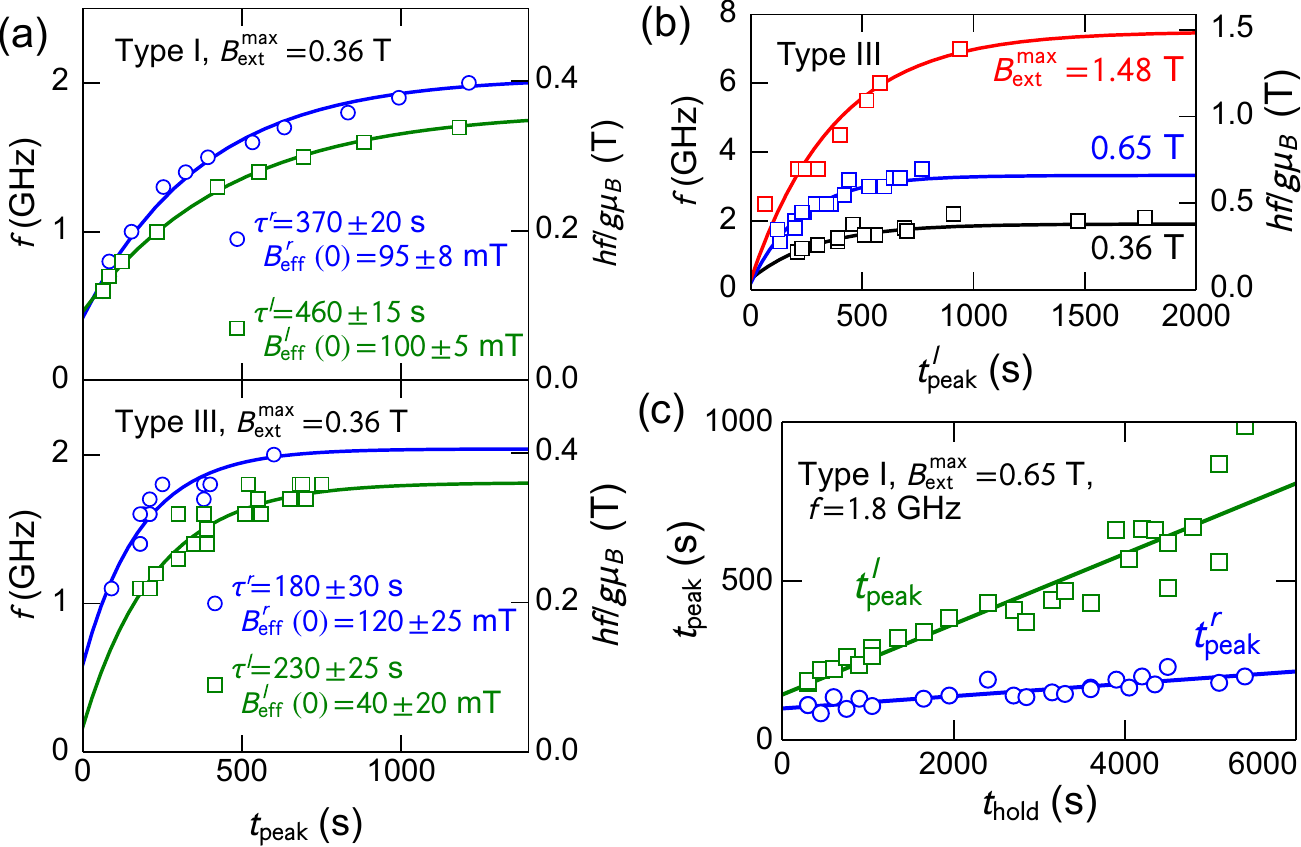}
\caption{ 
(a) EDSR resonance frequency versus time $f(t_\text{peak})$: \tr\ (blue, right dot) and \tl\ (green, left dot) for type I (III) curves in the upper (lower) panel. Lines are fits according to \eq{eqn:dnsp_decay}.
(b) Type III decay curves $f(\tl)$ for various \Bextmax.
(c) Slowing down the decay of the nuclear spin polarization:
$t_\text{peak}(t_\text{hold})$.
} 

\label{fig:dnsp_decay}
\end{figure}
To map out the decay of $\Bnr$ in the two dots, we present \tr\ and \tl\ (x-axis) of type I and III curves for various frequencies (y-axis) and $\Bextmax=0.36\,$T in \fig{fig:dnsp_decay}(a). The solid lines are theory curves, where we assume that \Bnr\ is antiparallel to $\mathbf{B}_\text{ext}$ (an assumption fully compatible with our data \cite{Petersen2013}) and decays exponentially, such that
\begin{eqnarray}
	hf = \left|g\mu_B\left[\Bextmax + B_\text{nm}^z + B_\text{nuc}^z(t=0)\exp(-t/\tau)\right]\right|\,,
	\label{eqn:dnsp_decay}
\end{eqnarray}
where we replaced vectors by their $z$-components (along the external field),
a good approximation for the data in \fig{fig:dnsp_decay}(a,b) \footnote{We expect deviations if the quantization axis deviates from the $z$-axis, which is never the case for our measurements taken at $hf \gg g\mu_B B_\text{nm}^\perp$.}.
In accordance with EDSR measurements on this sample \cite{Forster2015}, we use $g=-0.36$ and  $B_\text{nm}^z=55\,$mT vs.\ 10\,mT in the left vs.\ right dot. Note that $B_\text{nm}^z$ defines the long-time limit of the EDSR frequencies in each dot, and hence also $\tl-\tr$ in this limit. The knowledge of $\Bmr$ allows us to attribute the two EDSR resonances to the individual dots. We find that both, decay times and \Bnr\ generally differ from FP to FP and between the two dots, see \fig{fig:dnsp_decay}(a). In these measurements, decay times range from 3 minutes to almost 8 minutes, much longer than earlier findings in lateral dots \cite{Vink2009,Petersen2013}.

For both types I and III, our measurements show the build-up of a strong nuclear polarization which partially compensates $\Bext$ and $B_\text{nm}$ as evidenced by the small values of $\Beff$, see Fig.~\ref{fig:dnsp_decay}(a). FP I is characterized by a nuclear spin polarization which tends to equalize \Beff\ in the two dots: From the small current, we already concluded that $\Delta\Beff^\perp\simeq0$. From our EDSR measurements, we find $\Bnz= -(270\pm5)$\,mT vs.\ {$-(320\pm9)$\,mT} in the left vs.\ right dot, corresponding to  $\Delta\Beff^\parallel=(-5\pm10)\,$mT, hence $\Beffd\simeq0$. This complete compensation of $\mathbf B_\text{ext}+\Bmr$ is surprising as the EDSR measurement is performed at dot positions shifted by $\sim10\,$nm compared to where they are during polarization build-up. It is a first indication that nuclear spin polarization diffuses outside the dots. For FP III we find $\Bnz= -(340\pm25)$\,mT vs.\ $-(290\pm20)$\,mT in the left vs.\ right dot, corresponding to $\Delta\Beff^\parallel = (80 \pm 30)\,$mT (with statistical errors from the fit procedure).
Furthermore, FPs II and IV have similarly large leakage currents as FP III, indicating that these three FPs have in common a relatively large \Beffd\ at finite \Bext. Unfortunately, we have not been able to reliably capture EDSR current peaks following DNSP traces II and IV as these resonances turned out to be too unstable.

In \fig{fig:dnsp_decay}(b), we compare decay curves of the later occurring current maximum corresponding to the left dot for FP III and three different values of \Bextmax. We find equal decay times for relatively small \Bextmax, namely $\tau=230\pm20$\,s for $\Bextmax=0.36$\,T and $0.65$\,T, but a longer $\tau=400\pm60$\,s for $1.48$\,T. This points to a stabilization mechanism of the nuclear spin polarization inside the dots for longer sweeps.

To further explore the long decay times observed in \fig{fig:dnsp_decay}(a) and the stabilization mechanism evidenced in  \fig{fig:dnsp_decay}(b) we continued polarizing with $V=1\,$mV at a fixed $\Bextmax=0.65\,$T and measured \tr\ and \tl\ as a function of the additional polarizing time $t_\text{hold}$. As shown in \fig{fig:dnsp_decay}(c) we find that both current maxima occur later in proportion to the increase of $t_\text{hold}$. Similar lifetime enhancement of  \Bnr\ was previously observed in bulk GaAs \cite{Paget1982} and self-assembled quantum dots \cite{Makhonin2008,Latta2011}. Our long polarization lifetimes of many minutes for large $t_\text{hold}$ and the linear dependence of $t_\text{peak}$ on $t_\text{hold}$ confirm a significant polarization of the surroundings of the dots. The large and widely spread magnetic field inhomogeneity caused by our two nanomagnets further increases the lifetime by slowing nuclear spin diffusion via flip-flop processes. The latter are restricted by energy conservation but the nuclear spin state energies depend on the local magnetic and  electric (and strain) fields, all having varying gradients throughout the DQD. In our system, the gradient of $B_\text{nm}^z$ is around $0.2$--$1\,$mT/nm \cite{Forster2015}. The ensuing difference in Zeeman energy between closest homonuclear atoms is several times larger than their nuclear spin dipole-dipole coupling and of the same order as the  Knight-field gradient caused by the inhomogeneous electron wave function and estimated to reduce diffusion coefficients by a factor 2--10 \cite{Deng2005}. Since the magnetic field gradient extends far beyond the dots \cite{Forster2015}, it also facilitates the polarization of their surroundings. Increasing $t_\text{hold}$ from $0$ to $100$ minutes causes \tl\ to be delayed by more than a factor four while \tr\ is only increased by 70\,\%. We ascribe this difference to asymmetries in geometry and magnetic field in the two dots. 

In summary, we have dynamically polarized nuclear spins at a DQD and combined transport spectroscopy with electric-dipole-induced spin resonance to study the polarization and decay dynamics of nuclear spins. Measuring the leakage current through the DQD in Pauli-spin blockade, we find a remarkably complex current behavior during magnetic field sweeps. The statistical re-occurrence  of four patterns in $I(\Bext)$ establishes the existence of multiple fixed points, one of which is always occupied as long as dynamical polarization is maintained. Our EDSR measurements reveal long decay times of the nuclear spin polarization, its stability being enhanced by the strongly inhomogeneous magnetic field distribution generated by two single-domain nanomagnets in an extended area including the DQD. In addition, the EDSR measurements confirm that the individual FPs substantially  differ by their polarizations and dynamics of the nuclear spins. On one hand, our studies demonstrate that the existence of several FPs in dynamical nuclear spin polarization complicates the desired control of electron and nuclear spins in coupled quantum dots. On the other hand, our experiments present a salient advance in our understanding of the hyperfine induced dynamics in nanoelectronic circuits and brings us closer towards the desired fine control of the nuclear spins, important for quantum information applications. 

We are grateful for financial support from the DFG via SFB-631 and the
Cluster of Excellence \emph{Nanosystems Initiative Munich}. G.G. acknowledges
support by the Spanish Ministerio de Econom\'{\i}a y Competitividad
(MINECO) through the Project FIS2014-55987-P. S.\,L.\ acknowledges
support via a Heisenberg fellowship of the DFG.

%%%%%%%%%%%%%%%%%%%%%%%%%%%%%%%%%%%%%%%%%%%%%%%%%%%%%%%%%%%%%%%%%%%%%

%\bibliographystyle{prsty}
\bibliography{dnsp_esr,DNPreferencesGG}

\end{document}